\documentclass[pre,twocolumn,a4paper,noshowpacs]{revtex4}
\usepackage{ifpdf}
\usepackage[usenames]{color}
\usepackage{graphicx}
\usepackage{amssymb}
\usepackage{amsmath}
\usepackage{subfigure}

\begin{document}

\title{Reinforcement and inference in cross-situational word learning}

\author{Paulo F. C. Tilles and Jos\'e F. Fontanari}
\affiliation{Instituto de F\'{\i}sica de S\~ao Carlos,
  Universidade de S\~ao Paulo,
  Caixa Postal 369, 13560-970 S\~ao Carlos, S\~ao Paulo, Brazil}

\begin{abstract}
Cross-situational word learning is based on the notion that a learner can determine the referent of a word by finding something in common 
across many observed uses of that word. Here we propose an adaptive learning  algorithm that contains  a parameter that
controls the strength of the reinforcement applied to associations between concurrent words and referents, and 
a parameter that regulates inference, which includes built-in biases, such as   mutual exclusivity, and information of  past  learning events.
By adjusting  these parameters so  that the model predictions   agree with  data from representative experiments  on cross-situational word learning,
we  were able to explain the  learning strategies adopted by the participants of those experiments in terms of
a trade-off between reinforcement and inference. These strategies  can vary wildly  depending on the conditions of the
experiments. For instance, for   fast mapping  experiments  (i.e., the correct referent could, in principle, be inferred  in a
single observation) inference is prevalent, whereas for  segregated
contextual diversity experiments (i.e., the referents are separated in groups and are exhibited with members of their 
groups only) reinforcement is predominant. Other experiments are explained with more  balanced doses of reinforcement 
and inference.
\end{abstract}


\maketitle

\section{Introduction}\label{sec:intro}

A  desirable goal of  a  psychological theory  is to offer explanations grounded on elementary  principles  to the  
data available from psychology experiments
\citep{Newell_94}.
Although most of these quantitative psychological data is related to mental chronometry,  recent explorations on the human performance  to acquire an artificial lexicon in controlled laboratory conditions have paved the way  to the understanding of the learning strategies
humans use to infer a word-object mapping
\citep{Yu_07,Kachergis_09,Smith_11,Kachergis_12,Yu_12}. These experiments are based on the cross-situational word-learning paradigm
which  avers that  a learner can determine the meaning of a word by finding something in	common across all observed uses of that word 
\citep{Pinker_90,Gleitman_90}. In that sense, learning takes place through the statistical sampling of the contexts in which a word appears 
in accord with the classical associationist stance of  Hume and
Locke that the mechanism of word learning is sensitivity to covariation - if two events occur at the same time, they become associated
\citep{Bloom_00}.

In a typical cross-situational word-learning experiment,  participants are  exposed repeatedly to  multiple unfamiliar objects  concomitantly with
multiple spoken pseudo-words, such  that a word and its correct referent (object) always appear together on a learning trial.  Different trials
exhibiting distinct word-object pairs  will eventually allow the disambiguation of the word-object associations and the learning of
the correct mapping \citep{Yu_07}. It should be pointed out, however, that this scenario does not describe the actual word learning process by children
even in the unambiguous situation where the single novel object is followed by the utterance of its corresponding pseudo-word. In fact,   young children will
only make the connection between the object and the word provided they have a reason to believe that they are in presence of  an act of naming and for
this the speaker has to be present \citep{Baldwin_96,Bloom_00,Waxman_09}. Adults could learn those associations either because they  were previously
instructed by the experimenter that they would be learning which words go with which objects or because they could infer that the disembodied voice is
an act of naming by a concealed person. Nevertheless, the sort of statistical learning
assumed in the cross-situational word-learning  scenario may play a key role in  the development of pidgin \citep{Fontanari_11}  and
it is likely the sole option for the autonomous learning of a lexicon by robots \citep{Cangelosi_07,Steels_11}.

In order to learn a word-object mapping within the cross-situational word-learning scenario the learner should be able to
(i) recall at least a fraction of the word-object pairings that appeared in the learning trials, (ii) register both co-occurrences and non-co-occurrences
of words and objects and (iii) apply the mutual exclusivity principle which favors the association of novel words to novel objects \citep{Markman_88}.
Although a variety of rule-based and associative learning algorithms has been proposed to model cross-situational word learning 
\citep{Siskind_96,Frank_09,NN_09,Blythe_10,Kachergis_12,Tilles_12a,Reisenauer_13}, none of them explicitly take into account  processes 
 that regulate all three above-mentioned criteria. 

In this paper we  offer  an adaptive learning  algorithm that comprises two parameters which regulate  the
associative reinforcement of pairings between concurrent words and objects, and the non-associative 
inference process that  handles built-in biases (e.g., mutual exclusivity) as well as information of  past  learning events. 
By setting the values of these parameters so as to 
fit a  representative selection of  experimental  data presented in \citet{Kachergis_09,Kachergis_12} we are
able to identify and explain the  learning strategies adopted by the participants of those experiments in terms of
a trade-off between reinforcement and inference.

\section{Cross-situational  learning scenario}\label{sec:scen}

We assume  there are $N$ objects $o_1, \ldots, o_N$, $N$ words $w_1, \ldots, w_N$ and a one-to-one mapping between words and objects
represented by the set $\Gamma = \left\{ \left ( w_1, o_1 \right), \ldots, \left ( w_N, o_N \right) \right \}$. At each learning trial, $C$
word-object pairs are selected from $\Gamma$  and presented to the learner without providing any clue on which  word goes with which object.
For instance, pictures of the $C$ objects are displayed in a slide while  $C$ pseudo-words are spoken sequentially 
such that their spatial and temporal arrangements  do not give away the correct word-object associations \citep{Yu_07,Kachergis_09}.
We refer to the subset of words and their referents (objects)  presented to the learner in a learning trial as the context
$\Omega = \left\{w_{1}, o_{1}, w_{2}, o_{2},..., w_{C}, o_{C}\right\}$.
The context size $C$ is then a measure of the within-trial ambiguity, i.e., the number of co-occurring 
word-object pairs per learning trial.
The selection procedure  from the set $\Gamma$, which may favor some particular subsets of  word-object pairs, 
determines the different experimental setups discussed
in this paper. Although each individual trial is highly ambiguous, repetition  of trials with partially overlapping contexts 
 should in principle  allow the learning of the $N$ word-object associations. 

After the training stage is completed, which 
typically comprises about two dozen trials,  the learning accuracy is measured by instructing the learner to 
pick the object among the $N$ objects on display which the learner thinks is associated to a particular target word.  The test is repeated for
all $N$ words and the average learning accuracy calculated as the fraction of correct guesses \citep{Kachergis_09}.

This cross-situational learning scenario does not account for the presence of noise, such as the effect of out-of-context  words. 
This situation can be modeled by assuming that there is a certain  probability  (noise) that the  referent of one of the  spoken  words is not part of the context (so that word can be said to be out of context).  Although theoretical analysis shows that  there is a maximum noise intensity beyond which
statistical learning is  unattainable \citep{Tilles_12b}, as yet no experiment was carried out to verify the existence 
of  this  threshold phenomenon on
the learning  performance of human subjects.

\section{Model}\label{sec:model}

We model learning  as a change in the confidence with which the algorithm (or, for simplicity, the learner)  associates 
the word $w_i$ to an object  $o_j$ that
results from the observation and analysis  of the contexts presented in the  learning trials. More to the point, this confidence is represented by the probability $P_{t} \left( w_i,o_j\right)$
that $w_i$ is associated to $o_j$ at learning trial $t$. This probability is normalized such that $\sum_{o_j} P_{t} \left( w_i,o_j\right) = 1$ for
all $w_i$ and $t$, which then implies that  when the word $w_i$ is presented to the learner in the testing stage the learning  accuracy is
given simply by $P_{t} \left( w_i,o_i\right)$. In addition, we assume that  $P_{t} \left( w_i,o_j\right)$  contains  information 
presented  in the learning trials up to and including  trial $t$ only.

If at learning trial $t$ the learner observes the  context
 $\Omega_t = \left\{w_{1}, o_{1}, w_{2}, o_{2},..., w_{C}, o_{C}\right\}$ then it can
 infer  the existence of two other informative sets. First, the set of the words (and their referents) that appear for the first time at trial $t$,
which we denote by
$\tilde{\Omega}_t = \left\{\tilde{w}_{1}, \tilde{o}_{1}, \tilde{w}_{2}, \tilde{o}_{2},..., \tilde{w}_{\tilde{C}}, \tilde{o}_{\tilde{C}_t}\right\}$. 
Clearly, $  \tilde{\Omega}_t \subseteq \Omega_t$ and $ \tilde{C}_t \leq C$.
Second, the set of words (and their referents) that do not appear in $\Omega_t$ but that have already appeared in the previous trials,  
$\bar{\Omega}_t = \left\{\bar{w}_{1}, \bar{o}_{1},..., \bar{w}_{N_{t}-C}, \bar{o}_{N_{t}-C}\right\}$ where
$N_t$ is the total number of different  words  that appeared in contexts up to and including trial $t$. Clearly, 
$\bar{\Omega}_t \cap \Omega_t = \varnothing$. 
The update rule of the confidences $P_{t} \left( w_i,o_j\right)$  depends
on  which of these three sets the word $w_i$ and the object $o_j$ belong to (if $i \neq j$  they may  belong to different sets). In fact, 
our learning algorithm comprises
a parameter $\chi \in \left[0,1 \right]$ that measures the associative reinforcement capacity and applies only to 
known words that appear in the current context, and a parameter $\beta \in \left[0,1 \right]$ 
that measures the inference capacity and applies either to known words that do not appear in the current
context or to  new words in the current context.  Before the experiment begins ($t=0$) we set $P_0 \left(w_{i}, o_{j}\right) =0$ for all
words $w_{i}$ and objects $o_{j}$.  Next we describe how the confidences are updated following the sequential
presentation of contexts.

In the first trial ($t=1$) all words are new ($\tilde{C}_1 = N_1 = C$), so we set
\begin{equation}\label{P1}
P_1 \left(\tilde{w}_{i}, \tilde{o}_{j}\right) = \frac{1}{C}  
\end{equation}
for $\tilde{w}_{i},\tilde{o}_{j} \in \tilde{\Omega} = \Omega$. 
In the second or in an arbitrary trial $t$  we expect to
observe contexts exhibiting   both novel and repeated words. Novel words must go through an 
inference preprocessing stage before the reinforcement procedure can be applied to them. This is so because if $\tilde{w}_{i}$
appears for the first time at trial $t$ then $P_{t-1} \left(\tilde{w}_{i}, o_{j}\right) =0 $ for all objects $o_j$ and since the
reinforcement is proportional to $P_{t-1} \left(\tilde{w}_{i}, o_{j}\right)$ the confidences associated to $\tilde{w}_{i}$
would never be updated (see eq.\ (\ref{r}) and the explanation thereafter). Thus when a novel word $\tilde{w}_{i}$ appear at trial $t \geq 2$,
we redefine its confidence values at the previous trial (originally set to zero) as
\begin{eqnarray}
P_{t-1} \left(\tilde{w}_{i}, \tilde{o}_{j}\right)  & =  & \frac{\beta}{\tilde{C}_t} + \frac{1 - \beta}{N_{t-1} + \tilde{C}_t}, \label{new_1} \\
P_{t-1} \left(\tilde{w}_{i}, o_{j}\right) & =  & \frac{1 - \beta}{N_{t-1} +\tilde{C}_t},  \label{new_2}\\
P_{t-1} \left(\tilde{w}_{i}, \bar{o}_{j}\right) & = & \frac{1 - \beta}{N_{t-1} +\tilde{C}_t}.  \label{new_3}
\end{eqnarray}
On the one hand, setting the inference parameter $\beta$ to its maximum value $\beta =1 $  enforces the mutual exclusivity principle
which requires  that the new word $\tilde{w}_{i}$ be associated with equal probability  to the $\tilde{C}_t$  new objects $\tilde{o}_{j}$ in the current context. 
Hence in the case $\tilde{C}_t = 1$ the meaning of the new word would be inferred in a single presentation. On the other hand, for $\beta = 0$ the new word is associated with equal probability to all objects already seen up to an including
trial $t$, i.e., $N_t = N_{t-1} + \tilde{C}_t$. Intermediate values of $\beta$ describe a  situation of imperfect  inference. Note that using eqs.\ (\ref{new_1})-(\ref{new_3}) we can easily
verify that $\sum_{\tilde{o}_j} P_{t-1} \left(\tilde{w}_{i}, \tilde{o}_{j}\right) + \sum_{{o}_j} P_{t-1} \left(\tilde{w}_{i}, {o}_{j}\right) +
\sum_{\bar{o}_j} P_{t-1} \left(\tilde{w}_{i}, \bar{o}_{j}\right)  = 1$, in accord with the normalization constraint.

Now we can focus on the update rule of the confidence $P_t \left ( w_i, o_j \right )$ in the case both word $w_i$ and object $o_j$
appear in the context at trial $t$. The rule applies both to repeated and novel words, provided the confidences of the novel words
are preprocessed according to eqs.\ (\ref{new_1})-(\ref{new_3}). In order to fulfill automatically the normalization condition for word $w_i$, 
 the increase of the confidence $P_t \left ( w_i, o_j \right )$ with $o_j \in \Omega_t$  must be compensated by the decrease of the confidences
$P_t \left ( w_i, \bar{o}_j \right )$ with $\bar{o}_j \in \bar{\Omega}_t$.  This can be  implemented by  distributing evenly the total flux of probability out of the latter
confidences, i.e.,  $\sum_{\bar{o}_{j} \in \bar{\Omega}_t} P_{t-1} \left(w_{i}, \bar{o}_{j}\right)$,   over the confidences 
$P_t \left ( w_i, o_j \right )$ with $o_j \in \Omega_t$. Hence the net gain of confidence  on the association between
$w_i$ and $o_j$  is given by
\begin{equation}\label{r}
r_{t-1} \left(w_{i}, o_j \right) =  \chi P_{t-1} \left(w_{i}, o_{j}\right) \frac{\sum_{\bar{o}_{j} \in \bar{\Omega}_t} P_{t-1} \left(w_{i}, \bar{o}_{j}\right)}
{\sum_{o_{j} \in \Omega_t} P_{t-1} \left(w_{i}, o_{j}\right)}
\end{equation}
where, as mentioned before, the parameter $\chi \in \left [ 0, 1 \right ]$ measures the strength of the  reinforcement process. Note that
if both $o_j$ and $o_k$ appear in the context together with $w_i$ then the reinforcement procedure should not create any distinction between 
the  associations $\left ( w_i, o_j \right )$ and $\left ( w_i, o_k \right )$. This result is achieved provided that the ratio of the confidence gains equals
the ratio of the confidences before  reinforcement, i.e., 
$r_{t-1} \left(w_{i}, o_j \right)/r_{t-1} \left(w_{i}, o_k \right) = P_{t-1} \left(w_{i}, o_j \right)/P_{t-1} \left(w_{i}, o_k \right) $. This is the reason
that the reinforcement gain of a word-object association given by eq.\ (\ref{r}) is proportional to the previous confidence on that
association.
The total increase in the confidences between $w_i$ and the  objects that appear in the context, i.e. 
$\sum_{o_{j} \in \Omega_t} r_{t-1} \left(w_{i}, o_{j}\right)$, 
equals the product of  $\chi$  and  the
total decrease in the confidences between $w_i$ and the  objects that do not appear in the context, i.e.
$\sum_{\bar{o}_{j} \in \bar{\Omega}_t} P_{t-1} \left(w_{i}, \bar{o}_{j}\right)$. So for $\chi = 1$ the confidences associated to objects
absent from the context  are fully  transferred to the confidences associated  to
objects present in the context.  Lower values of $\chi$  allows us to control the flow of confidence  from  
objects in $\bar{\Omega}_t$ to  objects in $\Omega_t$.

Most importantly, in order to implement the reinforcement process the learner should be able
to use the information about the previous trials, which is condensed on the confidence values  $P_{t} \left(w_{i}, o_{j}\right)$. The 
efficiency on the usage of this information is  quantified by the word and trial dependent  quantity
 $\alpha_t \left (w_i \right ) \in \left [ 0, 1 \right ]$ that allows for the interpolation between
the cases of perfect usage ($\alpha_t \left (w_i \right )  = 1$)  and complete neglect  ($\alpha_t \left (w_i \right )  = 0$) of the information
stored in the confidences $P_{t} \left(w_{i}, o_{j}\right)$. In particular, we assume that the greatest the certainty on the association
between word $w_i$ and  its  referent, the more efficiently that information is used by the learner.  A quantitative measure of the
uncertainty associated to the confidences regarding word $w_i$ is given by the entropy
\begin{equation}
H_{t} \left(w_{i} \right) = - \sum_{o_{j} \in  \Omega_t \cup \bar{\Omega}_t}
 P_{t} \left(w_{i},o_{j}\right) \log \left[ P_{t} \left(w_{i},o_{j}\right)\right]
\end{equation}
whose maximum  ($\log N_t$) is obtained by the uniform distribution 
$P_{t} \left(w_{i},o_{j}\right) = 1/N_t$ for all $o_j \in  \Omega_t \cup \bar{\Omega}_t$, and
whose minimum ($0$) by  $P_t \left(w_{i},o_{j}\right) = 1$ and  $P_t \left(w_{i},o_{k}\right) = 0$ for $o_k \neq o_j$.
So we define
\begin{equation}\label{alpha}
\alpha_{t} \left(w_{i} \right) = \alpha_{0} + \left(1 -\alpha_{0}\right) \left[1 -\frac{H_{t} \left( w_{i} \right)}{\log N_{t}}\right],
\end{equation}
where $\alpha_0 \in \left [ 0, 1 \right ] $ is a baseline efficiency corresponding to the maximum uncertainty about
the referent of a target word.

Finally, recalling that at
trial $t$ the learner has access to the sets $\Omega_t$, $\bar{\Omega}_t$ as well as  to the confidences at trial $t-1$ we  write
the update rule
\begin{eqnarray}\label{Pwo}
P_{t} \left( w_{i}, o_{j} \right)  &  = &  P_{t-1}  \left( w_{i}, o_{j} \right)  + \alpha_{t-1} \left (w_i \right )  r_{t-1} \left( w_{i}, o_j \right) \nonumber \\
&  &  \! \!  + \left [1 -\alpha_{t-1} \left (w_i \right )  \right] \left[ \frac{1}{N_{t}} - P_{t-1} \left( w_{i}, o_{j} \right) \right] 
\end{eqnarray}
for  $w_{i}, o_{j} \in \Omega_{t}$. Note that if  $\alpha_{t-1}  \left (w_i \right ) = 0$ the learner would associate word 
$w_i$   to all objects that have appeared up to and including trial $t$ with equal probability. This situation happens
only if $\alpha_0 = 0$ and if there is complete uncertainty about the referent of word $w_i$.

Now we consider the update rule for the confidence $P_t \left ( w_i, \bar{o}_j \right )$ in the case that word $w_i$ appears in the
context at trial $t$ but object $\bar{o}_j$ does not. (We recall that  object $\bar{o}_j$ must have appeared in some previous trial.)
According to the reasoning that led to eq.\ (\ref{r})  this confidence  must decrease by the amount $\chi P_{t-1} \left ( w_i, \bar{o}_j \right )$ and so,
taking into account the information efficiency factor,
we  obtain
\begin{eqnarray} \label{Pwbo}
P_{t} \left( w_{i}, \bar{o}_{j} \right)  &  = &  P_{t-1}  \left( w_{i}, \bar{o}_{j} \right)  - \alpha_{t-1} \left (w_i \right )  \chi  P_{t-1}  \left( w_{i}, \bar{o}_{j} \right) \nonumber \\
&  &   + \left[1 -\alpha_{t-1} \left (w_i \right ) \right] \left[ \frac{1}{N_{t}} - P_{t-1} \left( w_{i}, \bar{o}_{j} \right) \right] 
\end{eqnarray}
which can be easily seen to satisfy the normalization 
\begin{equation}
 \sum_{o_j \in \Omega_t} P_{t} \left( w_{i}, o_{j} \right)  + 
\sum_{\bar{o}_j \in \bar{\Omega}_t} P_{t} \left( w_{i}, \bar{o}_{j} \right) = 1 .
\end{equation}

We focus now on the update rule for the confidence $P_t \left ( \bar{w}_i, \bar{o}_j \right )$ with $\bar{w}_i, \bar{o}_j \in
\bar{\Omega}_t$, i.e., both the word $\bar{w}_i$  and the object $\bar{o}_j$ are absent from the context shown at trial $t$, but they have already
appeared, not necessarily together, in previous trials. A similar inference reasoning that led to the expressions for the preprocessing
of new words would allow the  learner to conclude that a word absent from the context should be associated to an object  that is also
absent  from it. In that sense, confidence should flow from the associations between 
$\bar{w}_i$ and objects  $o_{j} \in \Omega_t$ to the associations between $\bar{w}_i$ and objects  $\bar{o}_{j} \in \bar{\Omega}_t$. Hence,
ignoring the information efficiency factor for the moment, the
net gain to confidence $P_{t} \left(\bar{w}_{i}, \bar{o}_{j}\right)$ is given by
\begin{equation}\label{rb}
	\bar{r}_{t-1} \left(\bar{w}_{i}, \bar{o}_j \right) =  \beta  P_{t-1} \left(\bar{w}_{i}, \bar{o}_{j}\right) \frac{\sum_{o_{j} \in \Omega_t} P_{t-1} 
\left( \bar{w}_{i}, o_{j}\right)}
	{\sum_{\bar{o}_{j} \in \bar{\Omega}_t} P_{t-1} \left(\bar{w}_{i}, \bar{o}_{j}\right)}.
\end{equation}
The direct proportionality of this gain to $P_{t-1} \left(\bar{w}_{i}, \bar{o}_{j}\right)$ can be justified by an argument  similar to
that used to justify eq.\ (\ref{r}) in the case of reinforcement. The information efficiency issue is also handled in a similar manner  
so the desired update rule reads
\begin{eqnarray} \label{Pbwbo}
P_{t} \!\left( \bar{w}_{i}, \bar{o}_{j} \right)  &  = &  P_{t-1}  \! \left( \bar{w}_{i}, \bar{o}_{j} \right)  + 
\alpha_{t-1} \! \left (\bar{w}_i \right ) \bar{r}_{t-1} \! \left( \bar{w}_{i}, \bar{o}_j \right) \nonumber \\
&  &  \! \! \! + \left[1 -\alpha_{t-1} \! \left (\bar{w}_i \right ) \right] \! \left[ \frac{1}{N_{t}} - P_{t-1} \! \left( \bar{w}_{i}, \bar{o}_{j} \right) \!\right] 
\end{eqnarray}
for  $\bar{w}_{i}, \bar{o}_{j} \in \bar{\Omega}_{t}$.  To
 ensure normalization  the confidence $P_{t} \left( \bar{w}_{i}, o_{j} \right)$
must decrease by an amount proportional to $\beta  P_{t-1} \left( \bar{w}_{i}, o_{j}\right)$  so that
\begin{eqnarray}\label{Pbwo} 
P_{t} \! \left( \bar{w}_{i}, o_{j} \right)  &  = &  P_{t-1}  \! \left( \bar{w}_{i}, o_{j} \right)  - \alpha_{t-1} \! \left (\bar{w}_i \right )  
\beta  P_{t-1}  \! \left( \bar{w}_{i}, o_{j} \right) \nonumber \\
&  &  \! \!  + \left[1 -\alpha_{t-1} \! \left (\bar{w}_i \right ) \right]  \! \left[ \frac{1}{N_{t}} - P_{t-1} \! \left( \bar{w}_{i}, o_{j} \right) \right] 
\end{eqnarray}
for $\bar{w}_i \in \bar{\Omega}_t$ and $o_j \in \Omega_t$.
 We can verify that  prescriptions (\ref{Pbwbo}) and  (\ref{Pbwo}) satisfy the normalization
\begin{equation}
 \sum_{\bar{o}_j \in \bar{\Omega}_t} P_{t} \left( \bar{w}_{i}, \bar{o}_{j} \right)  + 
\sum_{o_j \in \Omega_t} P_{t} \left( \bar{w}_{i}, o_{j} \right) = 1, 
\end{equation}
as expected.

In summary, before any trial ($t=0$) we set all confidence values to zero, i.e. $P_{0} \left( {w}_{i}, {o}_{j} \right) = 0$, and
fix the values of the parameters $\alpha_0$, $\chi$ and $\beta$.
In the first trial ($t=1$) we set the confidences of the words  and objects in $\Omega_1$ according to eq.\ (\ref{P1}),
so we have the values of $P_{1} \left( {w}_{i}, {o}_{j} \right)$ for ${w}_{i}, {o}_{j} \in \Omega_1$.
In the second trial, we separate the novel words  $\tilde{w}_i \in \tilde{\Omega}_2$ and reset 
$P_1\left( \tilde{w}_{i}, {o}_{j} \right)$  with $o_i \in \Omega_2 \cup \bar{\Omega}_2$ according to eqs.\ (\ref{new_1})-(\ref{new_3}).
Only then we calculate $\alpha_1 \left ( w_i \right ) $ with $w_i \in \Omega_1 \cup \tilde{\Omega}_2$  using eq.\ (\ref{alpha}).
The confidences at trial $t=2$ then follows from eqs.\ (\ref{Pwo}), (\ref{Pwbo}), (\ref{Pbwbo}) and (\ref{Pbwo}). As before, in the third trial
we separate the novel words $\tilde{w}_i \in \tilde{\Omega}_3$, reset $P_2\left( \tilde{w}_{i}, {o}_{j} \right)$  with $o_i \in \Omega_3 \cup \bar{\Omega}_3$ according to eqs.\ (\ref{new_1})-(\ref{new_3}), calculate $\alpha_2 \left ( w_i \right ) $ with $w_i \in \Omega_1 \cup \Omega_ 2 \cup 
\tilde{\Omega}_3$  using eq.\ (\ref{alpha}), and only then resume the evaluation of the confidences at trial $t=3$. This procedure
is repeated until the training stage is completed, say, at $t=t^*$. At this point, knowledge of  the confidence values
 $P_{t^*} \left ( w_i, o_j \right)$ allows us to answer any question posed in the  testing stage.

In the next section we evaluate the adequacy of this algorithm to describe a selection of cross-situational word-learning experiments
carried out on adult  subjects by \citet{Kachergis_09,Kachergis_12}.

\begin{figure*}[t]
\centering
\includegraphics[width=0.32\linewidth]{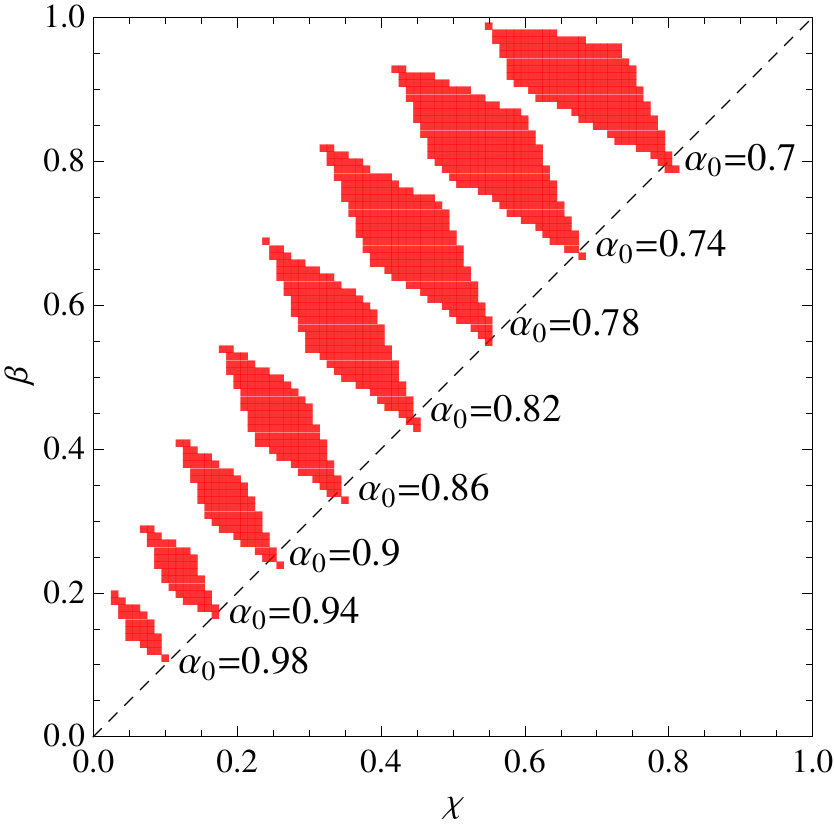}
\includegraphics[width=0.32\linewidth]{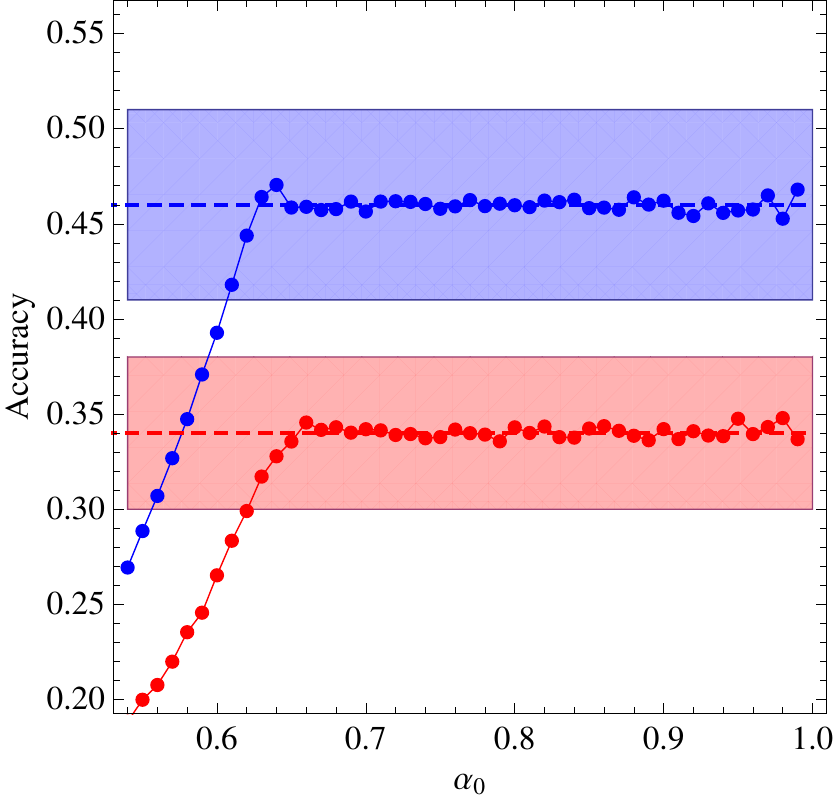}
\includegraphics[width=0.32\linewidth]{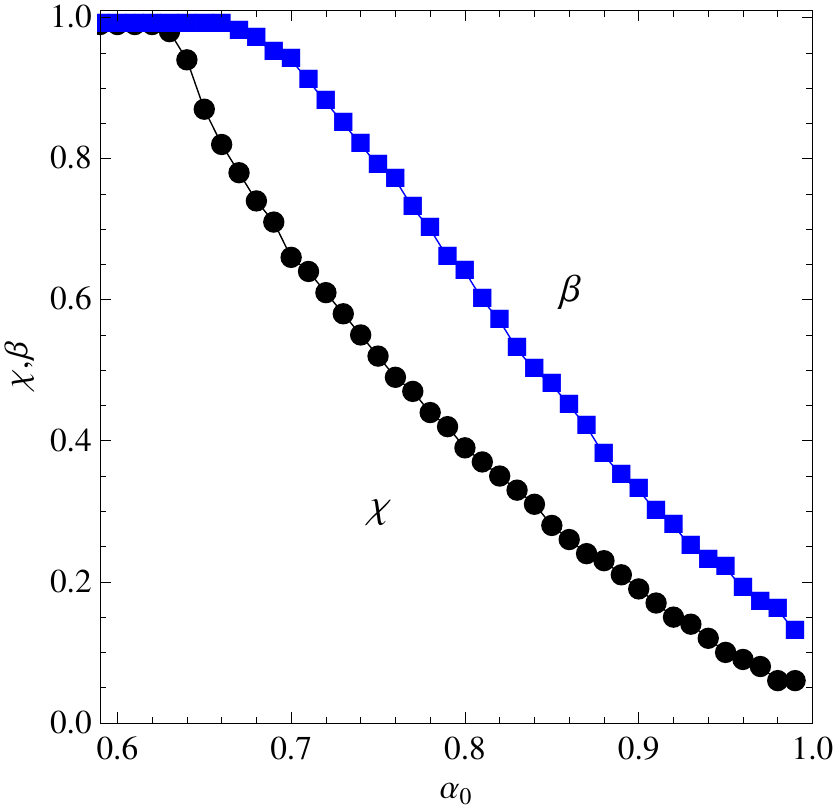}
\caption{Summary of the results for the two-frequency condition experiment. Left panel: Regions in the plane $\left ( \chi, \beta \right )$ where the algorithm fits the experimental data for fixed $\alpha_0$ as indicated in the figure.
Middle panel: Average accuracy for the best fit to the experimental results of \citet{Kachergis_09} 
represented by the broken horizontal lines  (means) and shaded regions around them (one standard deviation). 
The blue symbols represent the accuracy for the group of words
sampled 9 times whereas the red symbols represent the accuracy for the words sampled 3 times.
Right panel: Parameters $\chi$ and $\beta$ corresponding to the best fit shown in the middle panel. The other parameters are $N=18$ and $C=4$.}
\label{fig:1}
\end{figure*}

\begin{figure*}[t]
\centering
\includegraphics[width=0.32\linewidth]{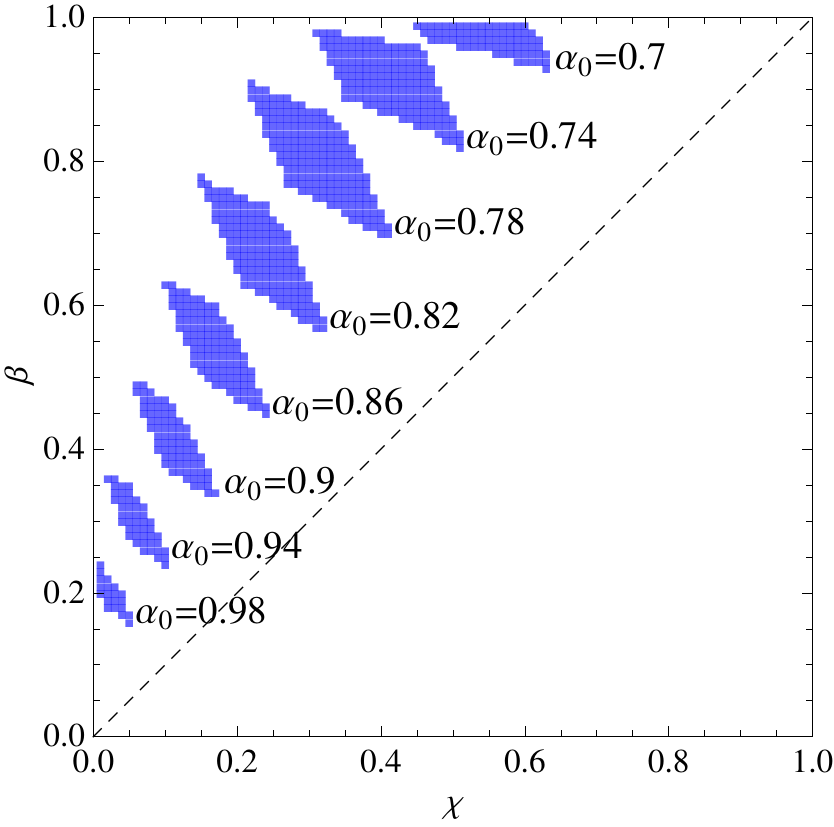}
\includegraphics[width=0.32\linewidth]{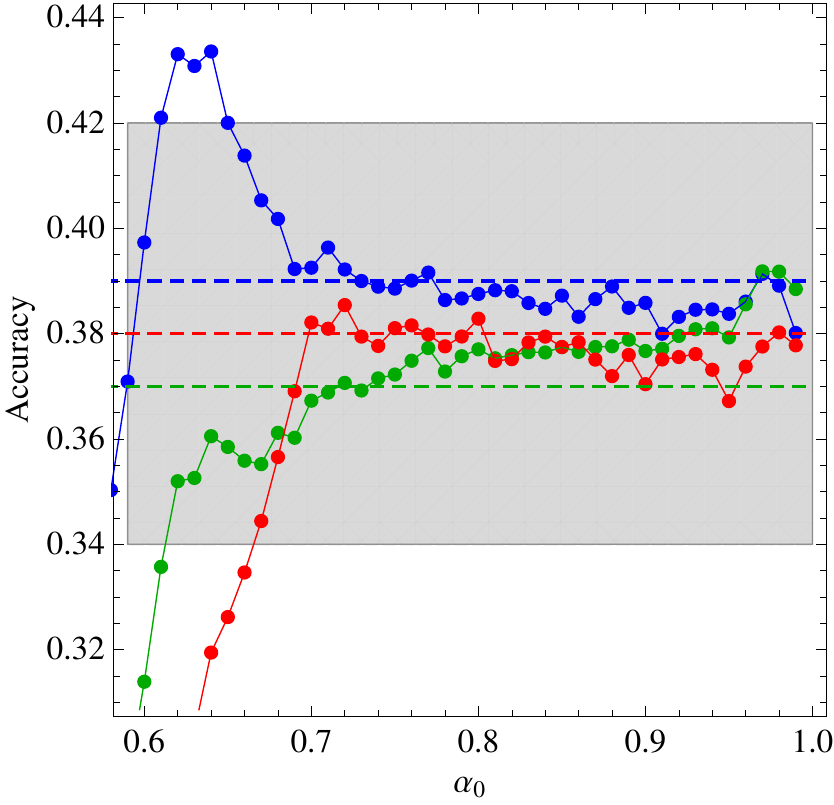}
\includegraphics[width=0.32\linewidth]{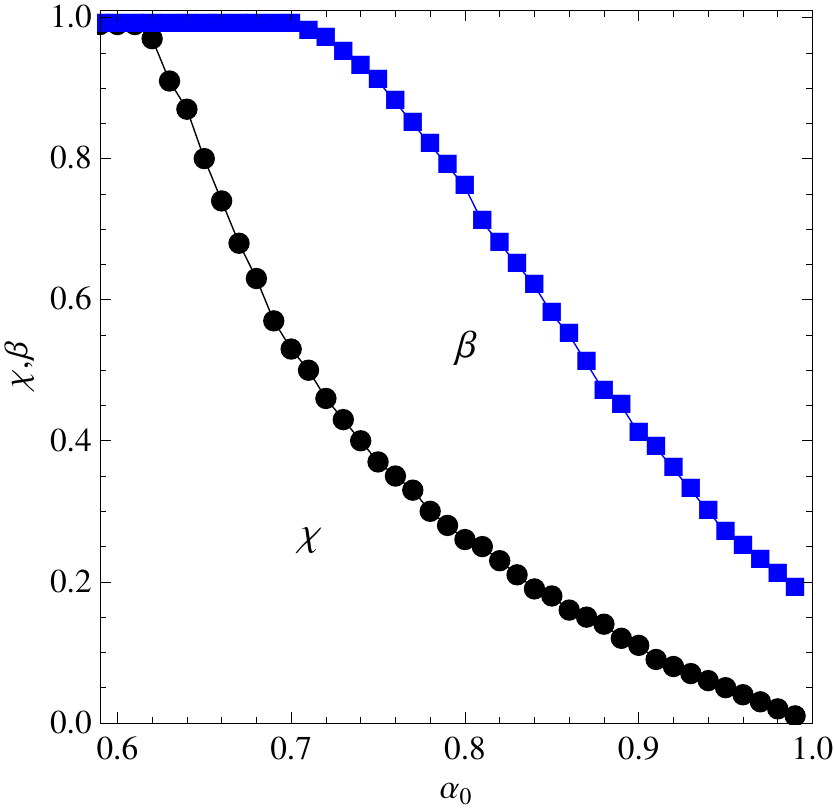}
\caption{Summary of the results for the three-frequency condition experiment. Left panel: Regions in the plane $\left ( \chi, \beta \right )$ where the algorithm fits the experimental data for fixed $\alpha_0$ as indicated in the figure.
Middle panel: Average accuracy for the best fit to the experimental results of \citet{Kachergis_09} represented by the broken horizontal lines  (means) and shaded regions around them (one standard deviation).
 The blue symbols represent the accuracy for the group of words
sampled 9 times,  the green symbols for the words sampled 6 times, and the red symbols for the words sampled 3 times.
Right panel: Parameters $\chi$ and $\beta$ corresponding to the best fit shown in the middle panel. The other parameters are $N=18$ and $C=4$.}
\label{fig:2}
\end{figure*}

\begin{figure*}[t]
\centering
\includegraphics[width=0.32\linewidth]{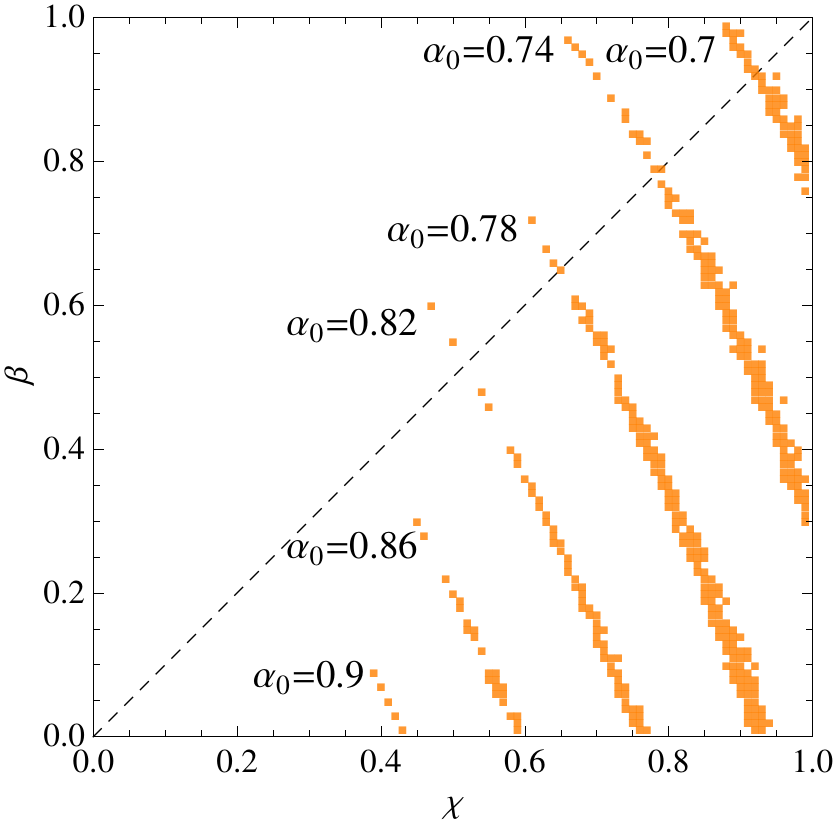}
\includegraphics[width=0.32\linewidth]{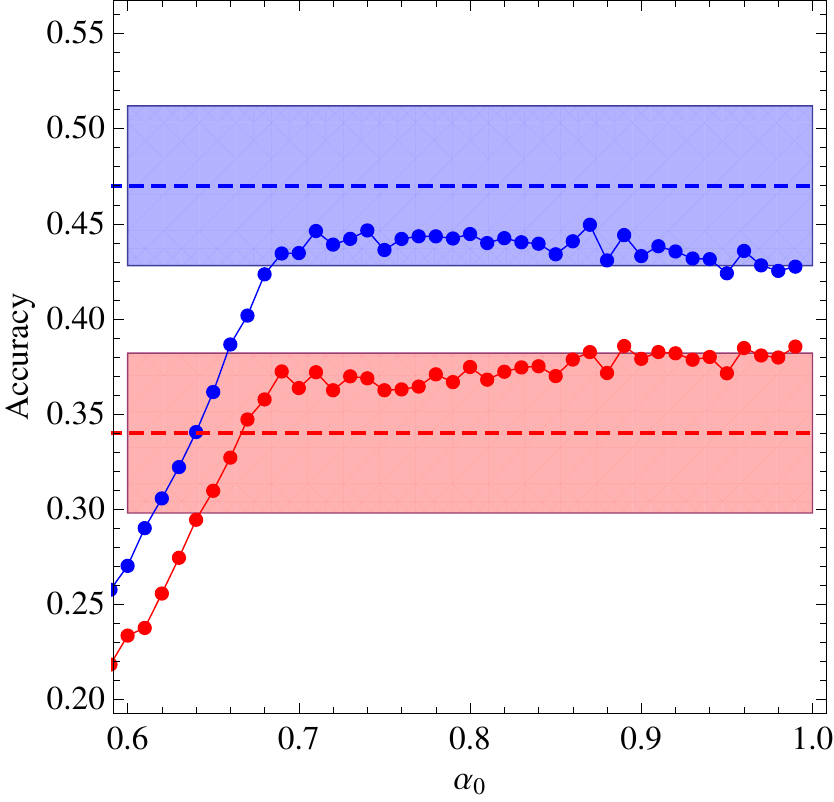}
\includegraphics[width=0.32\linewidth]{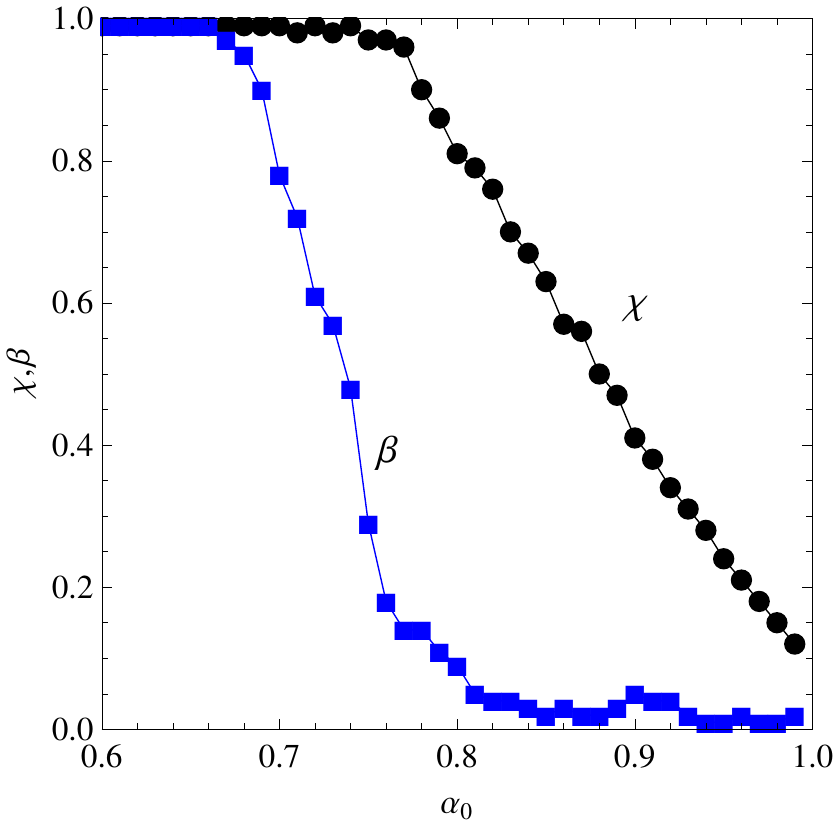}
\caption{
Summary of the results of the two-level contextual diversity experiment. Left panel: Regions in the plane $\left ( \chi, \beta \right )$ where the algorithm fits the experimental data for fixed $\alpha_0$ as indicated in the figure.
Middle panel: Average accuracy for the best fit to the experimental results of \citet{Kachergis_09}
represented by the broken horizontal lines  (means) and shaded regions around them (one standard deviation).
The blue symbols represent the accuracy for the group of words belonging to the 12-components subgroup and the red symbols for  the words belonging to the 6-components subgroup. All words are
repeated exactly 6 times during the $t^* = 27$ learning trials.
Right panel: Parameters $\chi$ and $\beta$ corresponding to the best fit shown in the middle panel. The other parameters are $N=18$ and $C=3$.}
\label{fig:3}
\end{figure*} 

\begin{figure*}[t]
\centering
\includegraphics[width=0.32\linewidth]{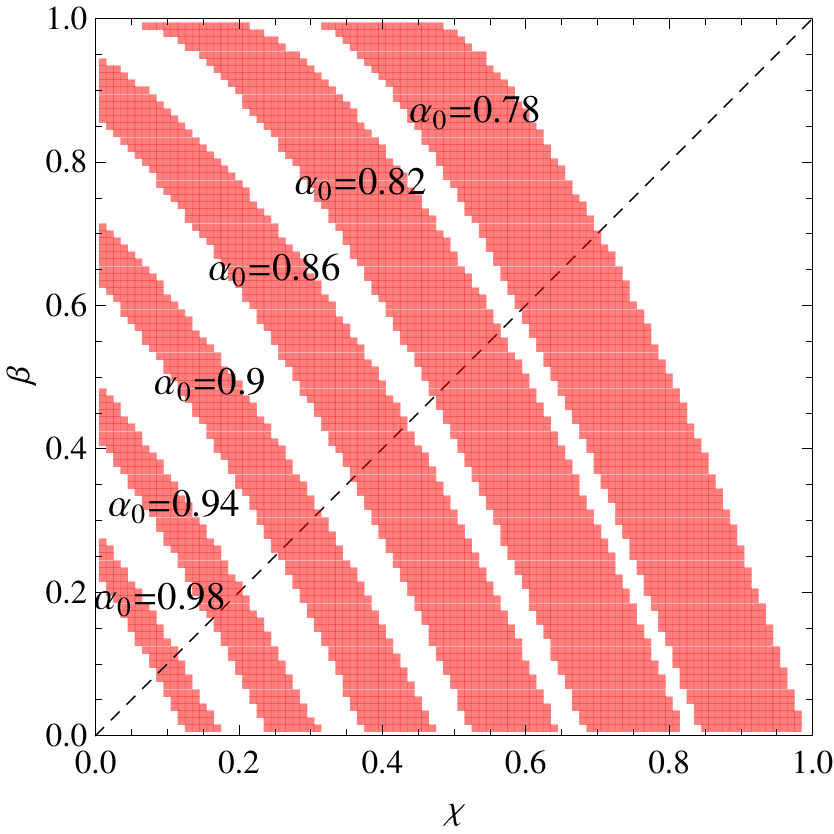}
\includegraphics[width=0.32\linewidth]{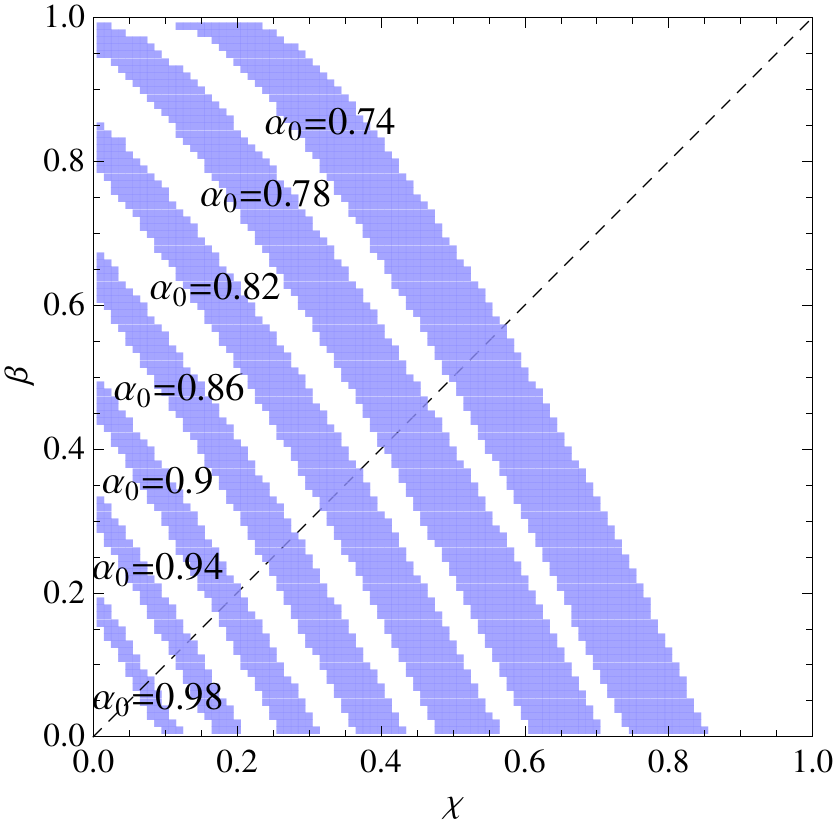}
\includegraphics[width=0.32\linewidth]{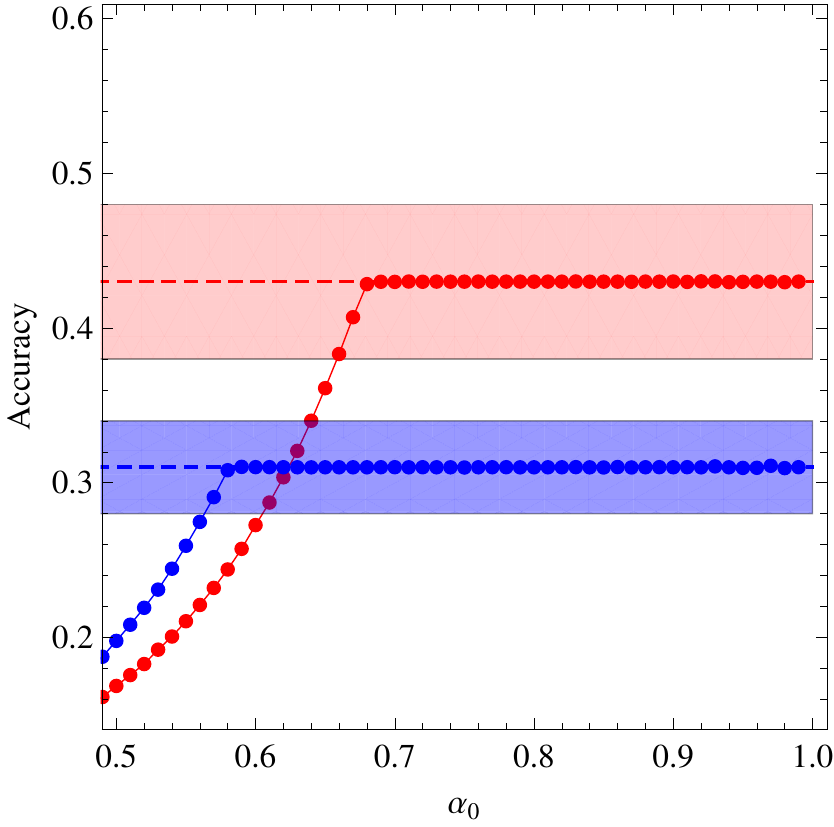}
\caption{Summary of the results of the experiments where all words co-occur without constraint and the $N=18$  words are
repeated exactly 6 times during the $t^* = 27$ learning trials. Left panel: Regions in the plane $\left ( \chi, \beta \right )$  where the algorithm fits the experimental data for fixed $\alpha_0$ and context size $C=3$. Middle panel: Same as the left panel but for context size $C=4$. Right panel: 
Average accuracy for the best fit to the experimental results of \citet{Kachergis_09} represented by the broken horizontal lines  (means) and shaded regions around them (one standard deviation).
The red symbols are for $C=3$ and the blue symbols  for $C=4$.}
\label{fig:4}
\end{figure*}

\section{Results}\label{sec:res}

The cross-situational word-learning experiments of \citet{Kachergis_09,Kachergis_12} aimed to understand how   
word sampling frequency (i.e., number of trials a word appears), contextual diversity (i.e., the co-occurrence of distinct words or groups of words in the learning trials),  within-trial ambiguity (i.e., the
context size $C$), and fast-mapping of novel words affect
the learning performance of adult subjects. In this section we compare the performance of the algorithm described in the previous section with the performance of
adult subjects reported  in \citet{Kachergis_09,Kachergis_12}.  In particular,
once   the conditions of the training stage are specified, we   carry out $10^4$ runs of  our algorithm for fixed values
of the three parameters $\alpha_0$, $\beta$, $\chi$, and  then calculate  the average accuracy  at  trial $t=t^*$ over all
those runs for that parameter setting. Since the algorithm is deterministic, what changes in each run is the composition of
the contexts at each learning trial.
As our goal is to model the results of the experiments, we search the space of parameters to find the
setting such that the performance of the algorithm matches that of humans within the error bars (i.e., one standard deviation) of the experiments.

\subsection{Word sampling frequency}\label{sub:freq}

In these  experiments the   number of words (and objects) is $N = 18$  and the training stage totals $t^* = 27$ learning trials, with each trial
comprising the presentation of  4 words together with their referents ($C=4$). Following  \citet{Kachergis_09}, we investigate two conditions
which differ with respect to the number of times a word is exhibited in  the training stage.
In the two-frequency condition, the 18 words are divided into two subsets of 9 words each. The words in the first subset appear 9 times and those
in  the second only 3 times. In the three-frequency condition, the 18 words are divided into three subsets of 6 words each.  Words in the first subset appear 3 times, in the second, 6 times and in the third, 9 times. In these two conditions, the same word was not allowed to appear in two consecutive learning trials.

 \textbf{Figures \ref{fig:1}} and  \textbf{\ref{fig:2}} summarize our main results for the two-frequency and three-frequency conditions, respectively.
The left panels show the regions (shaded  areas) in the $\left ( \chi, \beta \right)$ plane for fixed $\alpha_0$ where the algorithm describes the
experimental data. The middle panels show the accuracy of the  best fit as function of the parameter $\alpha_0$ and the right panels exhibit
the values of $\chi$ and $\beta$ corresponding to that fit.
 The broken horizontal lines and the shaded zones around them represent the means and standard deviations of the results of experiments carried out with 33 adult subjects \citep{Kachergis_09}. 

It is
interesting that although the  words sampled more frequently are learned best in the two-frequency condition as expected, this advantage practically
disappears in the three-frequency condition in which case all words  are  learned at equal levels  within the experimental error. Note that
the average accuracy for the words sampled 3 times is  actually greater than the accuracy for the words sampled 6 times, but this inversion is not
statistically significant, although, most surprisingly, the algorithm does  reproduce it 
for $\alpha_0 \in \left [0.7, 0.8 \right]$.  According to \citet{Kachergis_09}, the reason for the observed sampling frequency insensitivity might be because
the high-frequency words are learned quickly and once they are learned subsequent trials containing those words will exhibit an effectively smaller within-trial ambiguity. In this vein, the inversion could be explained if by chance the words less frequently sampled  were generally paired with the highly
sampled words. Thus  contextual diversity seems to  play a key role in cross-situational word learning.


\subsection{Contextual diversity and within-trial ambiguity}\label{sub:div}

In the first experiment aiming to probe the role of contextual diversity in the cross-situational learning, the 18 words were divided in 
two groups of 6 and 12 words each, and the contexts of size $C=3$ were formed with words  belonging to the same group only. 
 Since the sampling 
frequency was fixed to 6 repetitions for each word, those words belonging to the more numerous group are exposed
to a larger contextual diversity (i.e., the variety of different words with which a given word appear in the course of
the training stage). The results  summarized in \textbf{Figure \ref{fig:3}}  indicate clearly that contextual
diversity enhances the learning accuracy. Perhaps more telling is the finding that incorrect responses are
largely due to misassignments to referents whose words belong to the same group of the test word. In particular, \citet{Kachergis_09}
found that this type of error accounts for  56\% of incorrect answers when the test word belongs to the 6-components subgroup and 
for 76\%  when it belongs to the 12-components subgroup. The corresponding statistics  for our algorithm with the optimal parameters set
at $\alpha_0 = 0.9$ are  43\%  and 70\%, respectively. The region in the space of parameters where the model can be said to describe 
the experimental data is greatly reduced in this experiment and even the best fit is barely within the error bars. It is interesting that,
contrasting with the previous experiments,
in this case the reinforcement procedure seems to play the more important role in the performance of the algorithm.

The effect of the context size or within-trial ambiguity is addressed by the experiment summarized in \textbf{Figure \ref{fig:4}},
 which is similar to the previous experiment, except that the words  that compose the context are chosen uniformly from the entire repertoire of $N=18$ 
words. Two context sizes are considered, namely,  $C=3$ and $C=4$.  In both cases, there is a large selection of parameter values that  
explain the experimental data, yielding results indistinguishable from the experimental average accuracies. This is the reason we do not exhibit
a graph akin to those shown in the right panels of the previous figures. Since a perfect fitting can be obtained
both for $\chi > \beta$  and for $ \chi < \beta$, this experiment is uninformative with respect to these two abilities. As expected,
increase of the within-trial ambiguity difficilitate learning.  In addition, the (experimental) results for $C=3$ yield  a learning  accuracy value  that is
intermediary to those measured for the 6 and 12-components subgroups, which is in agreement with the 
conclusion that the increase of the contextual diversity enhances learning, since the mean number of different co-occurring words  is $4.0$
in the  6-components subgroup, $9.2$  in the 12-components subgroup and $8.8$ in the uniformly mixed situation \citep{Kachergis_09}.

\begin{figure*}[ht]
\centering
\includegraphics[width=0.32\linewidth]{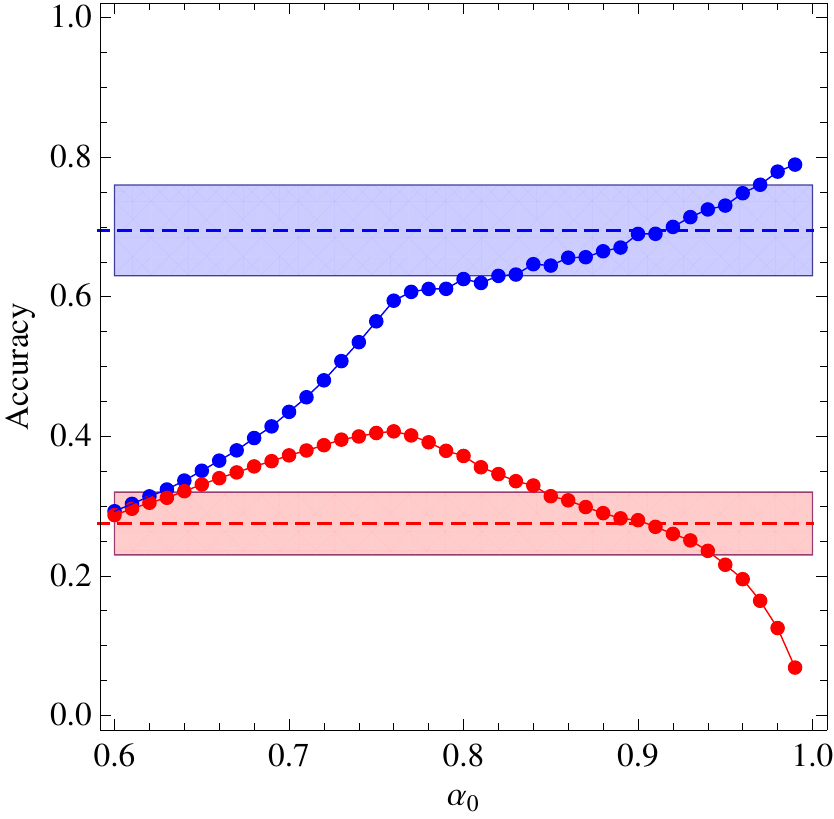}
\includegraphics[width=0.32\linewidth]{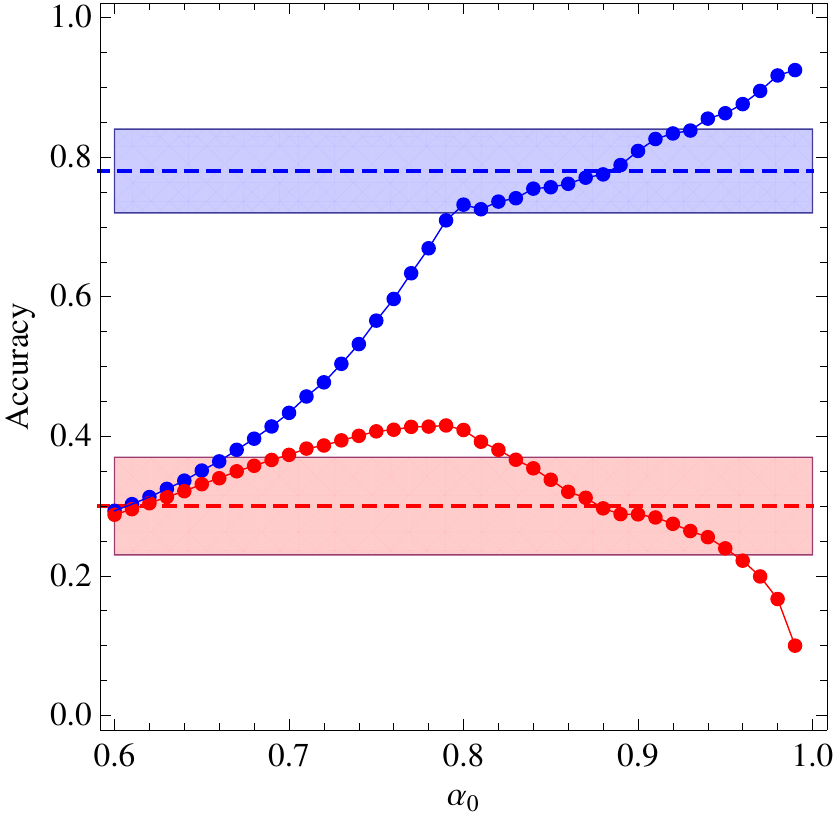}
\includegraphics[width=0.32\linewidth]{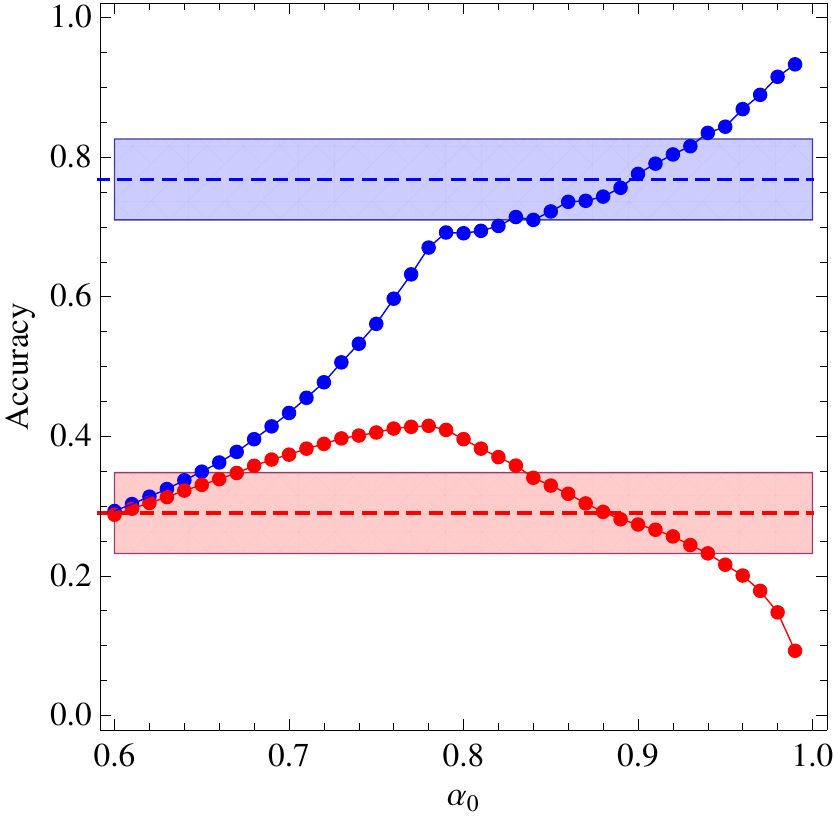}
\caption{Results of the experiments on mutual exclusivity  in the case the late phase of the training process
comprises 6 repetitions  of each word. The blue symbols represent the probability that the algorithm picks object $o_1$ as the referent
of word $w_1$ whereas the red symbols represent the probability it picks $o_7$. The broken horizontal lines and the shaded zones 
around them represent the experimental means and standard deviations \citep{Kachergis_12} represented by the broken horizontal lines  (means) and shaded regions around them (one standard deviation).
 The left panel shows the
results for 3 repetitions of $w_1$ in the early training phase, the   middle panel for 6 repetitions and the  right panel for
9 repetitions.  The  results correspond to the parameters $\chi$ and $\beta$ that  best fit the experimental data for fixed $\alpha_0$.  }
\label{fig:5}
\end{figure*}

\begin{figure*}[ht]
\centering
\includegraphics[width=0.32\linewidth]{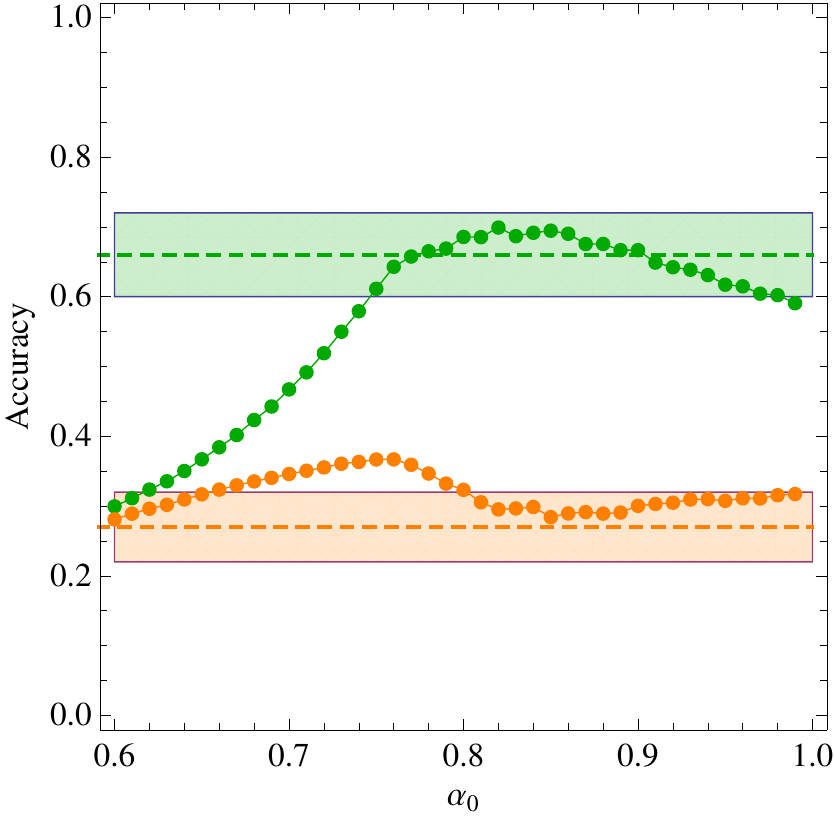}
\includegraphics[width=0.32\linewidth]{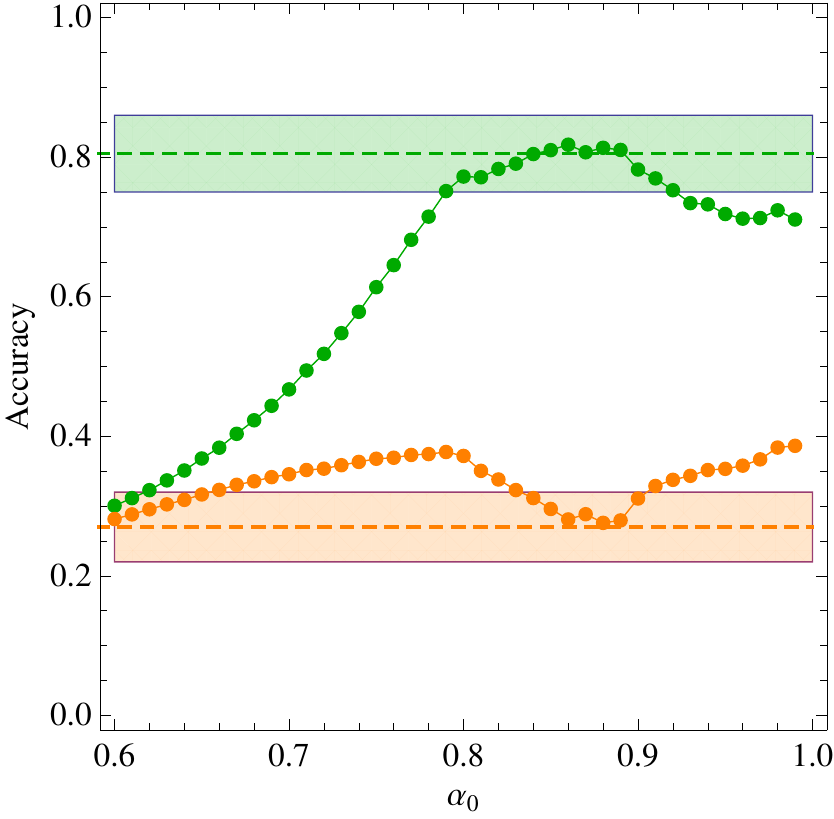}
\includegraphics[width=0.32\linewidth]{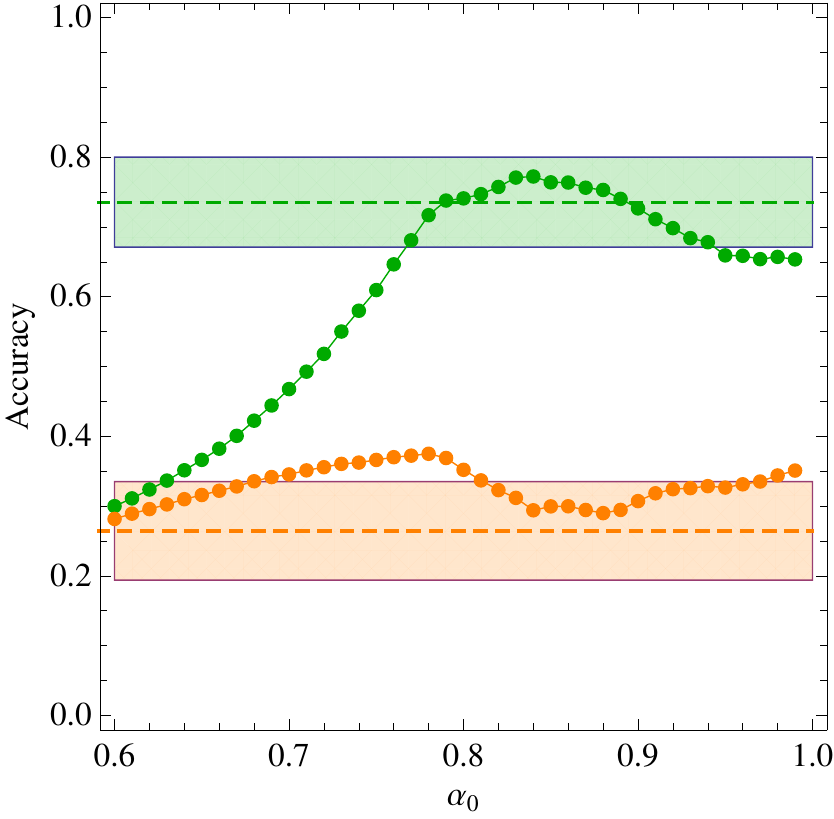}
\caption{Results of the experiments on mutual exclusivity  in the case the late phase of the training process
comprises 6 repetitions  of each word. The green symbols represent the probability that the algorithm picks object $o_7$ as the referent
of word $w_7$ whereas the orange symbols represent the probability it picks $o_1$. The broken horizontal lines and the shaded zones 
around them represent the experimental means and standard deviations \citep{Kachergis_12} represented by the broken horizontal lines  (means) and shaded regions around them (one standard deviation). The left panel shows the
results for 3 repetitions of $w_1$ in the early training phase, the   middle panel for 6 repetitions and the  right panel for
9 repetitions.  The  results correspond to the parameters $\chi$ and $\beta$ that  best fit the experimental data for fixed $\alpha_0$.}
\label{fig:6}
\end{figure*}

\subsection{Fast mapping}\label{sub:fast}

The experiments carried out by \citet{Kachergis_12} were designed to elicit participants' use of the mutual exclusivity principle
(i.e., the assumption of one-to-one mappings between words and referents) 
 and to test the flexibility of a learned word-object association when new evidence is provided in support to a many-to-many
mapping. To see how mutual exclusivity implies fast mapping assume  that a learner who  knows the
association $\left (w_1,o_1\right)$ is exposed to the context $\Omega = \{ w_1,o_1,w_2,o_2 \}$ in which the word $w_2$ (and its referent) appears
for the first time. Then it is clear that a mutual-exclusivity-biased learner would infer the association $\left (w_2,o_2 \right)$ in  this single
trial. However, a purely associative learner  would give equal weights to $o_1$ and $o_2$ if asked about the referent of $w_2$.

In the specific experiment we address in this section, $N=12$ words and their referents are split up into two groups of  6 words each, say
$A = \left \{ \left ( w_1, o_1 \right ), \ldots, \left ( w_6, o_6 \right ) \right \}$ and 
$B = \left \{ \left ( w_7, o_7 \right ), \ldots, \left ( w_{12}, o_{12} \right ) \right \}$.
The context size is set to $C=2$ and
the training stage is divided in two phases. In the early phase, only the words  belonging to group $A$ are presented and the
duration of this  phase is set such that each word is repeated 3, 6 or 9 times. In the late phase, the contexts consist of one
word belonging to $A$ and one belonging to $B$ forming fixed couples, i.e., whenever $w_i$  appears in a context, 
$w_{i+6}$, with $ i=1, \ldots, 6$, must appear too.  The duration of the late phase depends on the number of repetitions of each word that can be
3, 6, or 9 as in the early phase \citep{Kachergis_12}. The combinations of the  sampling frequencies   yield 9  different training conditions but here
we will consider only the case that the late phase comprises 6 repetitions of each word.

\begin{figure}[ht]
\centering
\includegraphics[width=19pc,angle=0]{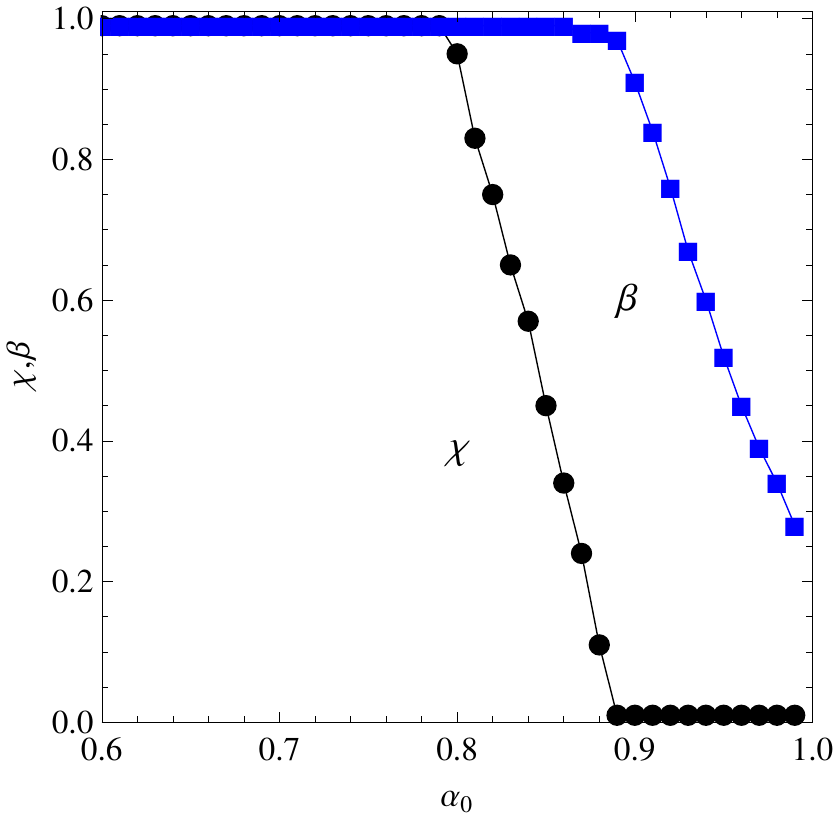}
\caption{Parameters $\chi$ (reinforcement strength) and $\beta$ (inference strength) corresponding to the best fit shown 
in Figures \ref{fig:5} and \ref{fig:6}  in the case word $w_1$  is repeated 6 times in the early training phase. 
}
\label{fig:7}
\end{figure}

The testing stage comprises the play of a single word, say $w_1$, and the display of 11  of the 12 trained objects \citep{Kachergis_12}.
Each word was tested twice with a time lag between the tests: once without its corresponding early object ($o_1$ in the case) and once without 
its late object ($o_7$ in the case). This procedure requires that we renormalize the confidences for each test. For instance,
in the case $o_1$ is left out of the display, the renormalization is
\begin{equation}
P_{t^*}' \left (w_1, o_j \right ) = P_{t^*} \left (w_1, o_j \right )/\sum_{o_k \neq o_1} P_{t^*} \left (w_1, o_k \right )
\end{equation}
with $j=2, \ldots, 12$ so that $\sum_{o_j \neq o_1} P_{t^*}' \left (w_1, o_j \right ) = 1$. 
Similarly, in the case $o_7$ is left out the renormalization becomes
\begin{equation}
P_{t^*}' \left (w_1, o_j \right ) = P_{t^*} \left (w_1, o_j \right )/\sum_{o_k \neq o_7} P_{t^*} \left (w_1, o_k \right )
\end{equation}
with $j=1, \ldots, 6,8,\ldots, 12$ so that $\sum_{o_j \neq o_7} P_{t^*}' \left (w_1, o_j \right ) = 1$.  We are interested  on the
(renormalized) confidences $P_{t^*}' \left (w_1, o_1 \right )$, $P_{t^*}' \left (w_1, o_7 \right )$, $P_{t^*}' \left (w_7, o_7 \right )$,
and $P_{t^*}' \left (w_7, o_1 \right )$, which are shown in \textbf{Figures \ref{fig:5}}  and \textbf{\ref{fig:6}} for the conditions 
where words $w_i, i=1, \ldots,6$  are repeated 3 (left panel), 6 (middle panel), and 9 (right panel)  times in the early learning phase,
 and  the words  $w_i, i=1, \ldots,12$ are repeated 6 times in the late phase.
The figures exhibit the performance of the algorithm for the set of parameters $\chi$ and $\beta$ that fits best the experimental 
data of \citet{Kachergis_12} for fixed  $\alpha_0$. This optimum set is shown in \textbf{Figure \ref{fig:7}} for the 6 early repetition condition, which
is practically indistinguishable from the optima of the other two conditions.  The conditions with the different  word repetitions in the early phase
intended to produce distinct  confidences on  the learned  association $\left ( w_1, o_1 \right )$ before the onset of  the late phase in the training stage.
The  insensitivity of the results  to these conditions probably indicates that  association was  already learned well enough with 3 repetitions only.
Finally, we note that, though the testing stage focused on  words 
$w_1$ and $w_7$ only, all word pairs $w_i$ and $w_{i+6}$ with $i=1,\ldots,6$ are strictly equivalent since they appear the same
number of times  during the training stage.

The experimental results exhibited in \textbf{Figure \ref{fig:6}} offer  indirect evidence that the participants have 
resorted to  mutual exclusivity to produce their  word-object mappings. In fact, from the perspective of a purely
associative learner, word $w_7$ should be associated to objects $o_1$ or  $o_7$  only, but
since in the testing stage one of those objects was not displayed, such a learner would  surely select the
correct referent. However, the finding that $P_{t^*}' \left (w_7, o_7 \right )$ is considerably greater than
$P_{t^*}' \left (w_7, o_1 \right )$ (they should be equal for an associative learner) indicates that there
is a bias against the association $\left ( w_7, o_1 \right)$  which is motivated, perhaps, from the previous understanding 
that $o_1$ was the referent of word $w_1$.  In fact, a most remarkable  result  revealed by  \textbf{Figure \ref{fig:6}} is 
that $P_{t^*}' \left (w_7, o_7 \right ) < 1$.  Since word $w_7$ appeared only in the late phase context 
$\Omega = \left \{ w_1, o_1, w_7, o_7 \right \}$ and object $o_1$ was not displayed in the testing stage, we must conclude
that the participants produced spurious associations between words and objects that never appeared together in a context.
Our algorithm  accounts for these associations  through  eq.\  (\ref{new_3}) in the case of new words and, more
importantly,  through eqs.\ (\ref{Pwbo})  and (\ref{Pbwo}) due to the effect of the information efficiency
factor $\alpha_t \left ( w_i \right)$. The experimental data is well described only in the narrow range
$\alpha_0 \in \left [0.85, 0.9 \right ]$.

\section{Discussion}\label{sec:dis}

The chief purpose of this paper  is to understand and model the mental processes  used by human subjects  to produce their word-object 
mappings in the controlled cross-situational word-learning scenarios devised  by  \citet{Yu_07} and \citet{Kachergis_09,Kachergis_12}. In other words,
we seek to analyze the psychological phenomena involved in the production of those mappings. Accordingly, we assume that the
completion of that task requires the existence of  two  cognitive abilities, namely, the associative capacity to create and reinforce
associations between words and referents  that co-occur in a context, and the non-associative capacity to infer word-object associations
based on  previous  learning events,  which  accounts for the mutual exclusivity principle, among other things.  In order to regulate  
the  effectiveness of these two
capacities we introduce the parameters $\chi \in \left [ 0, 1 \right ]$, which yields the reinforcement strength, and $\beta \in \left [ 0, 1 \right ]$,
which determines the inference strength.  

In addition, since the reinforcement and 
inference processes require storage, use  and transmission of past and present information (coded mainly on the  values of the confidences
$P_t \left ( w_i, o_j \right )$) 
we  introduce a  word-dependent quantity $\alpha_t \left ( w_i \right ) \in \left [ 0, 1 \right ]$ which measures how efficiently the
information regarding word $w_i$ is   processed. In particular, the greater the certainty about the referent of
word $w_i$, the more efficiently the information regarding $w_i$ is processed and transmitted from trial to trial.
 However, there is a 
baseline efficiency  $\alpha_0 \in \left [ 0, 1 \right ] $ used to process words for which the uncertainty about
their referents is maximum. The adaptive expression  for $\alpha_t \left ( w_i \right )$  given in eq.\  (\ref{alpha}) seems
to be critical for  the fitting of the experimental data. In fact,  our first choice was to use  a constant information efficiency 
(i.e., $\alpha_t \left ( w_i \right ) = \alpha ~ \forall t, w_i$)  with which we were able to describe only the experiments 
summarized in \textbf{Figures \ref{fig:1}} and \textbf{\ref{fig:4}} (data not shown). Note that a consequence of
prescription (\ref{alpha}) is that once the referent of a word is learned with maximum confidence (i.e.,
 $P_t \left(w_{i},o_{j}\right) = 1$ and  $P_t \left(w_{i},o_{k}\right) = 0$ for $o_k \neq o_j$) it is never
forgotten.

The algorithm described in Section \ref{sec:model} comprises three free parameters $\chi$,  $\beta$ and $\alpha_0$ 
which  are adjusted so as to fit a representative selection of the experimental data presented in \citet{Kachergis_09,Kachergis_12}. 
A robust result from all experiments is that the baseline information
efficiency is in the range $0.7 < \alpha_0 < 1$. Actually, the fast mapping experiments  narrow  this interval  down to
$0.85 < \alpha_0 < 0.9$. This is a welcome result because we do not have a clear-cut interpretation for $\alpha_0$ 
-- it encompasses storage, processing and transmission of information -- and so the fact that this parameter does not vary much
for wildly distinct experimental settings is evidence that, whatever its meaning, it is not relevant to
explain the learning strategies used in the different experimental conditions. 
Fortunately, this is not the case for the
two other  parameters $\chi$ and $\beta$. 

For instance, in the fast mapping experiments  discussed in Subsection \ref{sub:fast}
the best fit of the experimental data is achieved for $\beta \approx 1$ indicating thus the extensive use of mutual exclusivity, and inference
in general, by the participants of those experiments. Moreover, in that case  the best fit corresponds to  a low (but nonzero) 
value of $\chi$, which is expected since for contexts that exhibit  two associations ($C=2$)  only, most of the disambiguations are likely to 
be achieved solely through  inference.  This contrasts  with the experiments on variable word sampling frequencies  discussed in Subsection \ref{sub:freq}, for which the best fit is obtained with intermediate values of $\beta$ and $\chi$ so the participants' use of reinforcement and inference was not 
too unbalanced. The contextual diversity experiment  of Subsection \ref{sub:div}, in which the words are segregated in two isolated
groups of 12 and 6 components, offers another extreme learning situation, since  the best fit  corresponds
to $\chi \approx 1$ and $\beta \approx 0$ in that case.  To understand this result,  first we  recall that most of the participants' errors were due to 
misassignments of referents belonging to the same group of the test word, and  those confidences were strengthened mainly by the
reinforcement process. Second, in contrast to the  inference process, which creates  and strengthens spurious intergroup associations
via eq.\   (\ref{Pbwbo}), the reinforcement process solely weakens those associations  via eq.\ (\ref{Pwbo}). Thus, considering the
learning conditions of the contextual diversity experiment it is no surprise that reinforcement  was  the participants' choice  strategy.

Our results agree with the findings of \citet{Smith_11}  that participants  use various learning strategies, which in
our case are  determined by the values of the parameters $\chi$ and $\beta$, depending on the specific conditions
of  the cross-situational word-learning experiment. In particular, in the case of low within-trial ambiguity those authors
found  that participants generally resorted to a rigorous eliminative approach to infer the correct word-object mapping. This
is exactly the conclusion we reached  in the analysis of the fast mapping experiment for which the within-trial ambiguity
takes the lowest possible value ($C=2$).

Although the adaptive learning algorithm presented in this paper  reproduced the performance of
adult  participants in cross-situational word-learning experiments  quite  successfully, the deterministic nature of the algorithm
hindered somewhat the psychological interpretation of  the information efficiency factor $\alpha_t \left ( w_i \right )$.
 In fact, 
not only learning and behavior are best described as  stochastic processes \citep{Atkinson_65} but also the
modeling of those processes  requires (and facilitates) a precise interpretation of the model parameters, since
they  are introduced in the model as transition probabilities. Work in that line is underway.


\section*{Acknowledgement}
The work of J.F.F. was supported in part by Conselho Nacional de
Desenvolvimento Cient{\'\i}fico e Tecnol{\'o}gico (CNPq)  and P.F.C.T. was supported by  
grant  \# 2011/11386-1, S\~ao Paulo Research Foundation (FAPESP).



%
%

\end{document}